# RAMA: A new agile AO bench for the telescope FEELINGS

Mattéo Pasinetti[a,b], Rodrigo Andres Muñoz Gomez[d], Benjamín Andres González Barraza[d], Francisco Oyarzún[a], Sylvain Cetre[e], Jean-François Sauvage[c,a], Axel Vincent[c], Julien Charton[f], Marie Laslandes[f], Roméo Roudeix[f], Nicolas Védrenne[c], Cyril Petit[c], Pierre-Louis Mayeur[c], Esteban Vera Rojas[d], Andrew Reeves[b], Perrine Lognoné[b], Morgan Gray[a], Thierry Fusco[c,a], and Benoit Neichel[a]

[a]Aix Marseille University, CNRS, CNES, LAM, Marseille, France
[b]Durham University, Durham, UK
[c]ONERA, DOTA, F-13661 Salon cx Air – France
[d]Pontificia Universidad Católica de Valparaíso, Valparaíso, Chile
[e]Wakea Consulting, Grenoble, France
[f]ALPAO, Montbonnot-Saint-Martin, France

## ABSTRACT

The ground-based observation of extended objects such as satellites suffer from severe atmospheric propagation constraints. Specifically, high tracking velocities and low-elevation lines of sight generate non-stationary turbulence alongside strong scintillation. Developing robust wavefront control strategies is therefore critical to maintain stable observations. In this context, we present RAMA, an adaptive optics (AO) testbench deployed on ONERA's 60 cm FEELINGS telescope. Designed as a pathfinder for future systems like the PROVIDENCE ground station, RAMA evaluates a visible, non-modulated Pyramid Wavefront Sensor (PWFS). The hardware baseline also includes two pupil-conjugated deformable mirrors (DM97 and DM192) driven by the DAO Real-Time Computer (RTC). Inheriting the modular and evolving philosophy of the PAPYRUS project, the bench provides a flexible environment to test new components on-sky and allows for direct comparisons between classical controllers and advanced, data-driven strategies. By implementing Convolutional Neural Networks (CNN) for phase reconstruction and Reinforcement Learning (RL) for loop control, RAMA aims to overcome the specific limitations associated with scintillation and extended-object observations. This paper details the opto-mechanical design, numerical simulations of the bench expected wavefont control performance, preliminary laboratory closed-loop results and the first on-sky optical coupling with the telescope.

## 1. INTRODUCTION

The ground-based observation of extended objects, natural or artificial (see Figure 1), are severely limited by atmospheric propagation. The fast angular velocities and low-elevation trajectories, which result in long atmospheric propagation paths, of these objects induce highly non-stationary turbulence and severe scintillation.[1–3] In addition, optical observations typically occur near sunrise or sunset, where rapid changes in both sky background brightness and target illumination create strong fluctuations in the Signal-to-Noise Ratio (SNR).[1,3] Operating from non-astronomical sites like the FEELINGS station further exacerbates these issues, requiring robust wavefront control strategies (see Figure 2)

Natural bodies, such as asteroids, typically exhibit slower apparent velocities and present a relatively uniform illumination.[4] In contrast, artificial targets like LEO satellites dictate a much harsher tracking regime. They are characterized by extreme angular velocities, complex proper motions, and highly non-uniform, rapidly fluctuating illumination signatures.[1,3] These dynamic variations severely complicate the wavefront sensing process.

Further author information:
Mattéo Pasinetti: E-mail: matteo.pasinetti@lam.fr

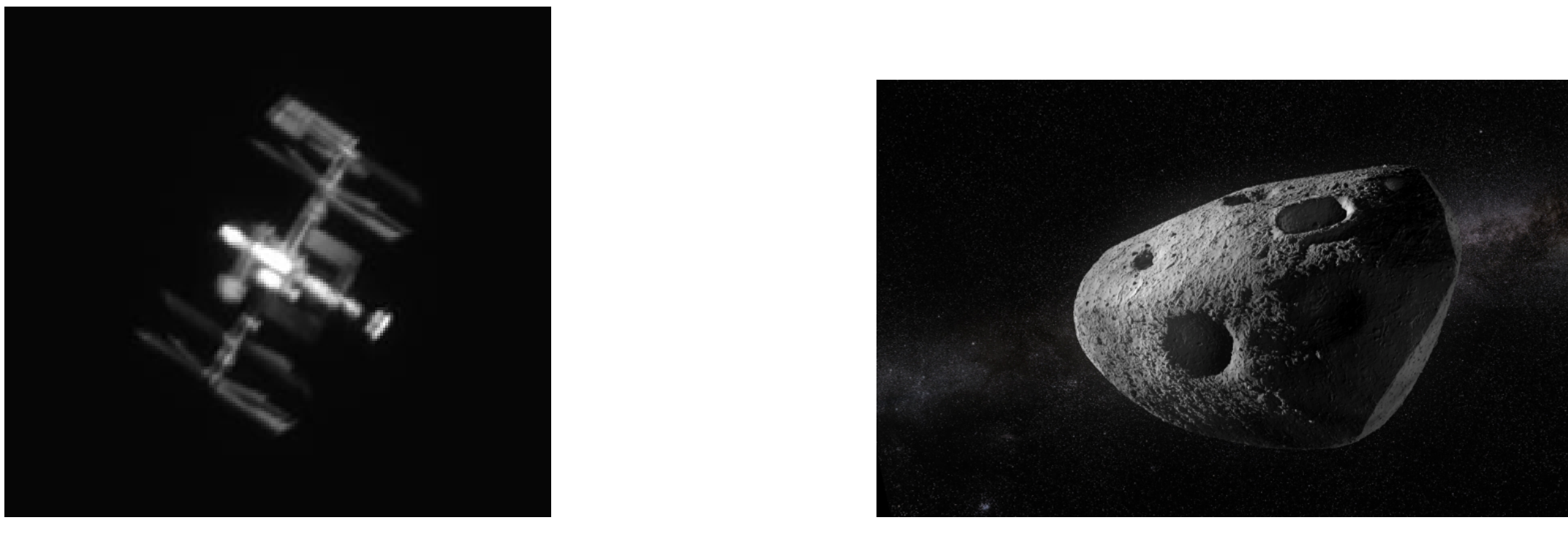

(a) International Space Station (b) Apophis asteroid

Figure 1: Example of type of extended objects that the bench could be used to observe.

While Shack-Hartmann Wavefront Sensors (SHWFS) remain the standard for extended targets observations,[1,5,6] the Pyramid Wavefront Sensor[7] (PWFS) offers high sensitivity and reduced pixel-readout requirements, fitting the extreme frame rates needed for fast-moving targets. Deploying the PWFS for extended objects, however, requires dedicated on-sky validation.

RAMA is an adaptive optics (AO) testbench installed on ONERA's 60 cm FEELINGS[8] telescope. Following the legacy of the PAPYRUS project[9,10] which was an AO bench placed at the 1.52m telescope of the Observatoire de Haute Provence, RAMA adopts an open and evolving testbench philosophy and will act as a versatile pathfinder for testing new components and algorithms on-sky before their integration into larger ground stations, such as the 2.5 m PROVIDENCE project.[11,12] The current configuration features a visible, non-modulated PWFS, two deformable mirrors (DMs) conjugated to the pupil, a visible science camera (Emergent IMX421LLJ), and the DAO Real-Time Computer[13,14] (RTC).

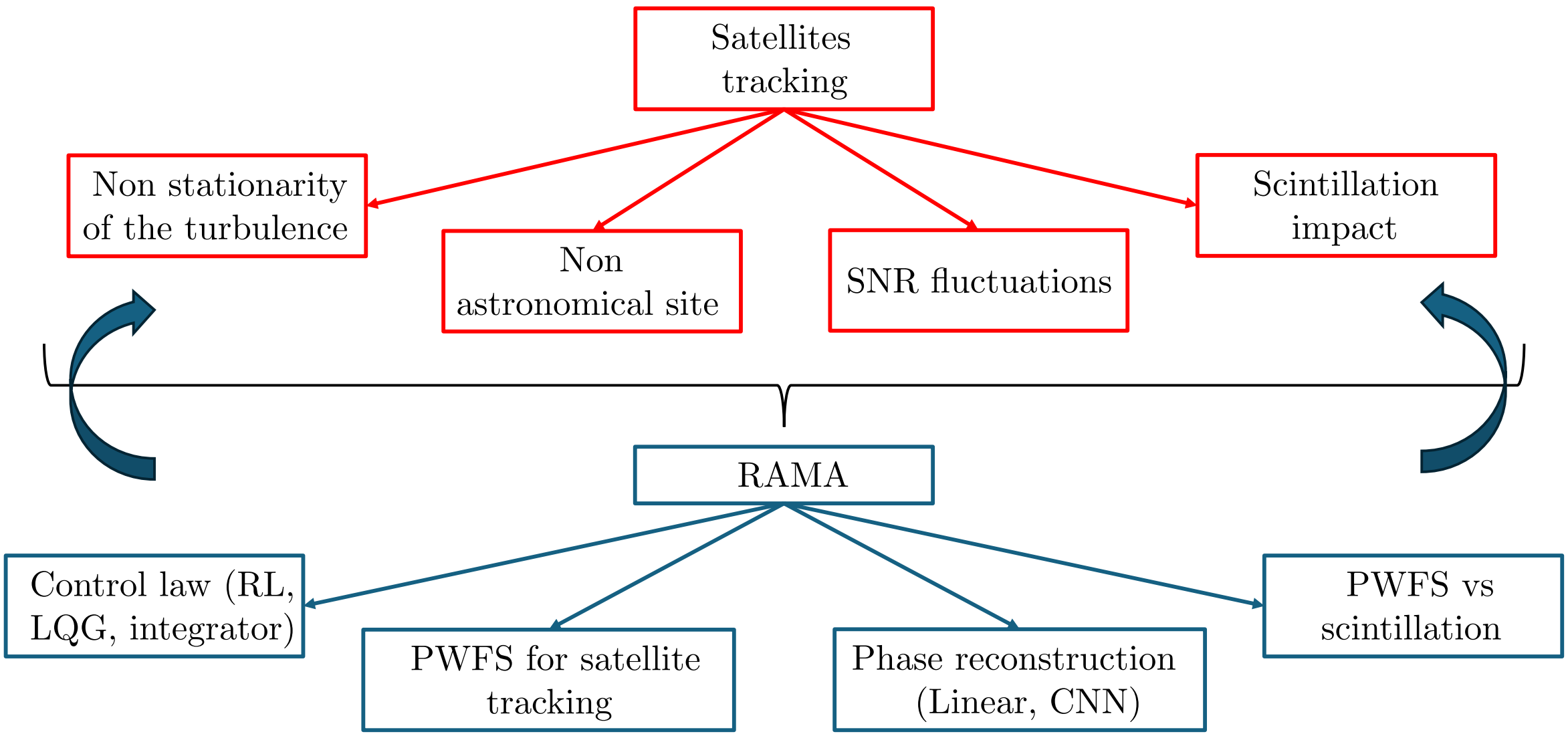

Figure 2: Scientific goals and challenges for the RAMA testbench, targeting satellite observations.

The setup aims to evaluate PWFS performances on extended targets and its resilience to strong scintillation. It also provides an experimental framework to benchmark conventional controllers against artificial intelligence approaches, including Convolutional Neural Networks[15–17] (CNN) for phase reconstruction and Reinforcement Learning[18–22] (RL) for predictive loop control.

This paper outlines the RAMA testbench. Section 2 details the opto-mechanical design and hardware con-straints. Section 3 covers numerical simulations of expected performances of the bench. Section 4 presents the project's current status and initial laboratory closed-loop results.

# 2. ADAPTIVE OPTICS DESIGN

## 2.1 Constraints and hardware baseline

The hardware and optics were taken from off-the-shelf components to ensure a quick integration of the bench and reduced costs. Wavefront correction is performed by two ALPAO deformable mirrors, a DM97 and a DM192 (currently a flat mirror), both conjugated to the telescope pupil. The entire hardware is interfaced with the DAO Real-Time Computer. On the current bench for now, the system will operate at loop frequencies up to 2000 Hz, a maximum rate constrained by the camera readout speeds in our current WFS branch optical design.

| Component | Model | Specifications | Role |
|---|---|---|---|
| Telescope | ONERA FEELINGS | 60 cm diameter | Light source |
| DM1 (DM192) | ALPAO | 15×15 actuators, pupil-conjugated (currently flat mirror) | Future woofer-tweeter control |
| DM2 (DM97) | ALPAO | 11×11 actuators, pupil-conjugated | Wavefront correction (primary) |
| PWFS | PAPYRUS pyramid | Rooftop: 25 $\mu$m, F/125 at apex, non-modulated | Wavefront sensing |
| Science camera | Emergent IMX421LLJ | 1944×1472 pixels, 4.5 $\mu$m/pixel, global shutter | PSF imaging / GSC |
| PWFS camera | Emergent IMX426LLJ | 816×624 pixels, 9.0 $\mu$m/pixel, global shutter | PWFS pupils imaging |
| RTC | DAO Real-Time Computer | 0–2000 Hz (upper limit from the camera readout due to the optical design) | Real-time control |

Table 1: Hardware parameters of the RAMA testbench.

The wavefront sensing approach, using a non-modulated visible Pyramid Wavefront Sensor (PWFS) and an Emergent IMX426LLJ camera, introduces a critical system limitation. Operating without modulation significantly increases the sensor's nonlinearity[23, 24] and its sensitivity to the physical imperfections of the component's apex (see figure 3a). Instead of a perfect point, the junction of the four faces forms a flat "rooftop" approximately 25 $\mu$m wide in our case. Without modulation, the core of the unmodulated Point Spread Function (PSF) rests statically on this imperfect region, causing an asymmetrical illumination of the Pyramid pupils degrading the phase reconstruction performances and therefore also the wavefront control capabilities of the system.

To evaluate the impact of this unmodulated configuration on closed-loop performance, numerical simulations were conducted by varying the ratio between the physical rooftop size and the optical PSF size. Figure 3b displays the resulting pupil intensity distributions. When the rooftop covers 25% of the PSF, the four pupil images remain well-defined and symmetrically illuminated. However, when this ratio increases to 34%, the intensity distribution starts to be highly asymmetric. The consequences on the adaptive optics loop are quantified in Figure 4, which plots the residual wavefront variance over time. The performance for rooftop sizes up to 25% is close to that of an ideal, defect-free pyramid. But exceeding this threshold (e.g., the 34% curve) results in difficulty closing the loop due to the saturation in the phase reconstruction that this asymmetry creates.

Consequently, to ensure stable non-modulated operation, the rooftop footprint must be strictly kept at or below 25% of the PSF size. Given the fixed physical rooftop dimension of 25 $\mu$m, the optical design is heavily constrained: the beam must be strongly focused to dilate the PSF, requiring a focal ratio of exactly F/125 at the pyramid apex.

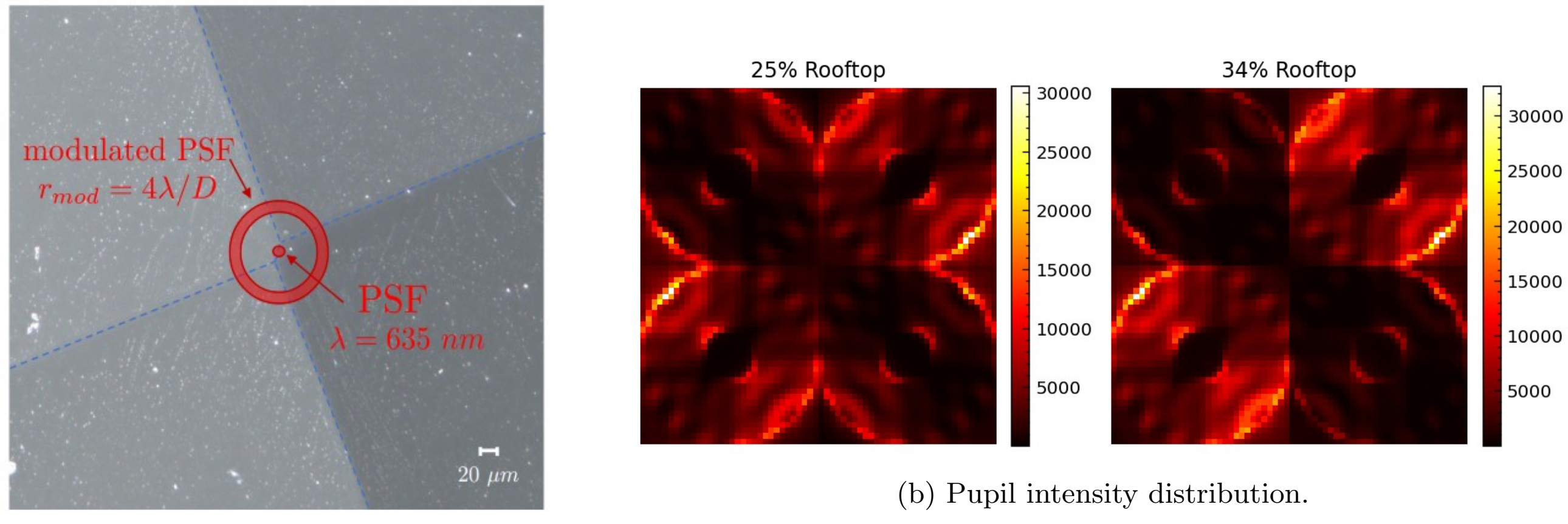

(a) Microscopic view.

(b) Pupil intensity distribution.

Figure 3: Physical pyramid rooftop (a) and its impact on the non-modulated PWFS pupil illumination (b).

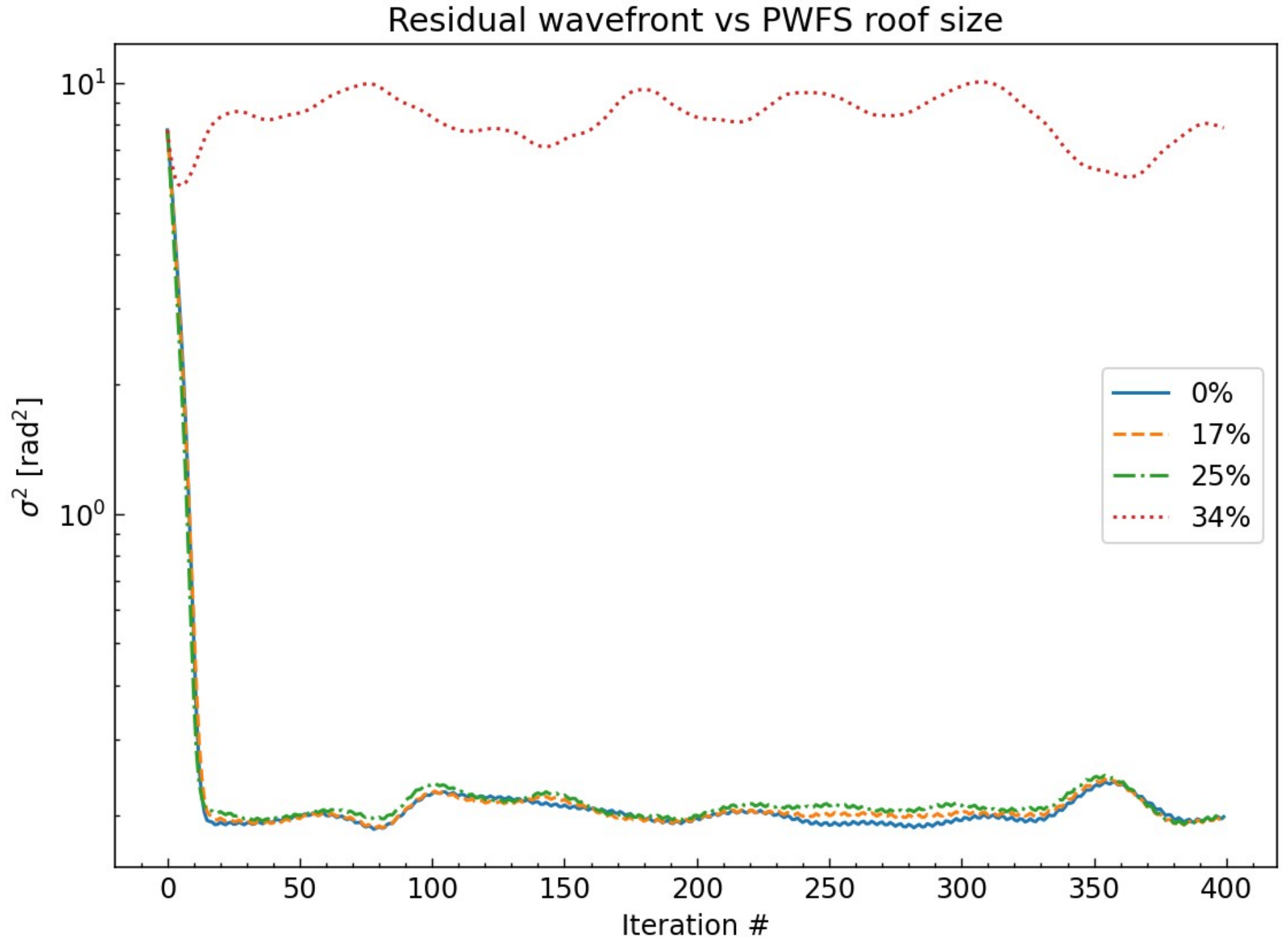

Figure 4: Residual wavefront variance over time for different rooftop sizes. Simulations demonstrate loop instability beyond a 25% ratio.

## 2.2 Optical design overview

The optical design for RAMA aims to satisfy the F/125 requirement at the PWFS top, in parallel with supporting extended-target observations under the constraint of a confined bench space.

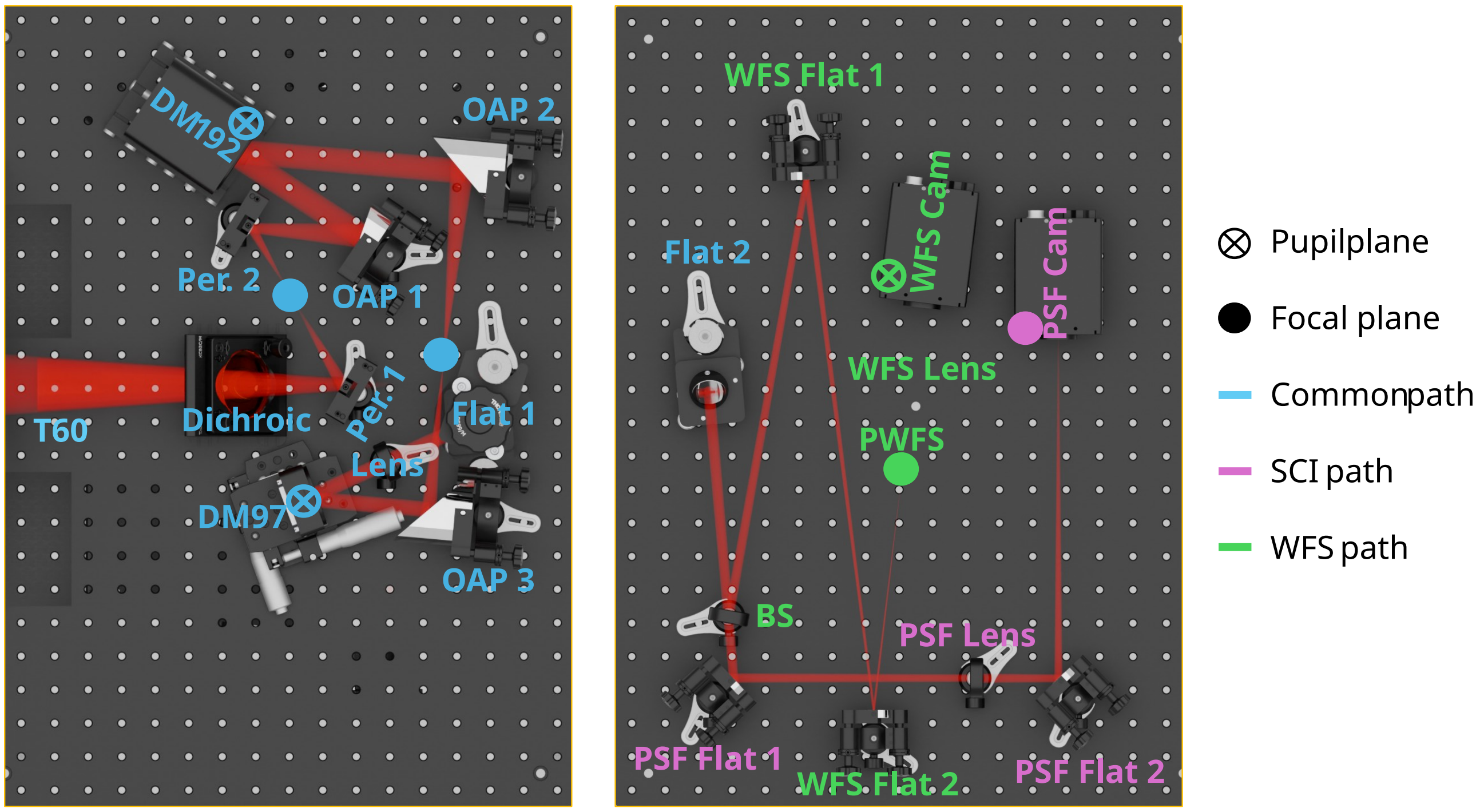

Figure 5: Optical layout of the RAMA testbench.

The common path begins with a periscope, consisting of two flat mirrors (Per. 1 and Per. 2), which redirects the incoming beam from the telescope (T60) after it passes through a visible dichroic (Dichroic). For standalone testing and interaction matrix calibration, a calibration source can be inserted at the focal plane between the two mirrors of the first periscope. Because the following off-axis parabolic mirror (OAP 1) collimates this beam, the internal propagation mimics on-sky observations. This first OAP conjugates the telescope pupil onto the first deformable mirror (DM192). A pair of off-axis parabolas (OAP 2 and OAP 3) then acts as an optical relay, reducing the beam diameter to 12 mm and conjugating the pupil onto the second deformable mirror (DM97). This specific design ensures that all critical pupil planes—the telescope, DM192, DM97, and the WFS camera—are strictly conjugated with one another.

To meet the focusing requirements of the non-modulated PWFS, a 1.5 m focal length (Lens) focuses the beam. A second periscope (Flat 1 and Flat 2) folds this converging beam down to the lower deck of the optical table, where a beam splitter (BS) divides the light into the wavefront sensing (WFS) and the scientific branches.

In the WFS branch, two flat mirrors (WFS Flat 1 and WFS Flat 2) provide the necessary propagation distance to form the focal point precisely at the apex of the pyramid (PWFS). This optical relay generates a highly focused F/125 focal ratio at the pyramid apex with a $\pm 15''$ FoV for the PWFS. A final 35 mm focal length lens (WFS Lens) then collimates the resulting four pupil images onto the WFS camera (WFS Cam).

The science branch utilizes two flat mirrors (PSF Flat 1 and PSF Flat 2) and a focusing lens (PSF Lens) to direct the beam onto the science camera (PSF Cam). Due to its shared focal plane conjugation with the pyramid, this camera functions as both a standard PSF imager and a Gain Scheduling Camera[25,26] (GSC) for the Optical Gain[25,26] (OG) estimation. Operating at an F/60.5 focal ratio with its native 4.5 $\mu$m pixels, the science camera provides a fine spatial sampling of 0.0256 arcsec/pixel. At the nominal wavelength of 635 nm, this corresponds to approximately 8.5 pixels per $\lambda/D$. This configuration gives a total field of view of 49.7 × 37.6 arcsec.

| Step | Component | Type | X (mm) | Y (mm) | Z (mm) | X tilt (°) | Beam type |
|---|---|---|---|---|---|---|---|
| 1 | T60 (Telescope input) | - | 0 | -300.00 | 0 | 90.00 | Convergent |
| 2 | Dichroic | Visible dichroic | 0 | 0 | -982.35 | 0 | Convergent |
| 3 | Per. 1 | Flat mirror | -1.57 | 0 | -1054.69 | 30.00 | Convergent |
| 4 | Per. 2 | Flat mirror | -1.57 | -116.91 | -987.19 | 32.00 | Divergent |
| 5 | OAP 1 | Off-axis parabola | -1.57 | -46.97 | -1037.23 | 34.15 | Collimated |
| 6 | DM192 | Flat mirror (for now) | -1.57 | -174.98 | -971.03 | 19.10 | Collimated |
| 7 | OAP 2 | Off-axis parabola | -1.57 | -0.94 | -1206.48 | 5.10 | Convergent |
| 8 | OAP 3 | Off-axis parabola | -1.57 | -3.20 | -1181.18 | -174.90 | Collimated |
| 9 | DM97 | Deformable mirror (11×11) | -1.57 | 85.36 | -1030.70 | 170.10 | Collimated |
| 10 | Lens | F=1500 mm | -1.57 | 58.00 | -1089.66 | 155.10 | Convergent |
| 11 | Flat 1 | Flat mirror | -1.57 | 25.54 | -1159.60 | 155.10 | Convergent |
| 12 | Flat 2 | Flat mirror | 298.43 | 25.54 | -1159.60 | 85.10 | Convergent |
| 13 | BS | Beam Splitter | 298.43 | -143.84 | -1145.09 | 93.10 | Convergent |
| 14 | WFS Flat 1 | Flat mirror | 298.43 | 179.98 | -1081.53 | 92.10 | Convergent |
| 15 | WFS Flat 2 | Flat mirror | 298.43 | -217.12 | -1033.51 | 90.10 | Convergent |
| 16 | PWFS | Pyramid | - | - | - | - | Focused (F/125) |
| 17 | WFS Lens | F=35 mm | - | - | - | - | Collimated |
| 18 | WFS Cam (WFS) | IMX426LLJ | 298.43 | 100.42 | -993.93 | 97.10 | Collimated |
| 19 | PSF Flat 1 | Flat mirror | 298.43 | -188.66 | -1141.41 | 133.10 | Convergent |
| 20 | PSF Flat 2 | Flat mirror | 298.43 | -193.35 | -898.20 | -133.90 | Convergent |
| 21 | PSF Lens | Lens | 298.43 | -192.13 | -961.45 | -178.90 | Convergent |
| 22 | PSF Cam (GSC) | IMX421LLJ | 298.43 | 76.60 | -893.00 | -88.90 | Focused |

Table 2: Optical design parameters of the RAMA testbench.

Finally, to address Non-Common Path Aberrations (NCPA) introduced by the beam splitter and the separated paths, the focal plane PSF will be directly optimized. The system will then be calibrated around this specific reference offset on the pyramid, enabling the non-modulated PWFS to close the loop while inherently compensating for the NCPA.

### 2.3 PAPYRUS tests for the calibration strategy

The calibration strategies proposed for RAMA were initially tested and validated on-sky using the PAPYRUS bench, providing a proven baseline from which to start. Building on this heritage, we propose a unified calibration strategy to robustly operate the RAMA non-modulated PWFS across two distinct observational scenarios: point sources and extended objects (see Figure 6).

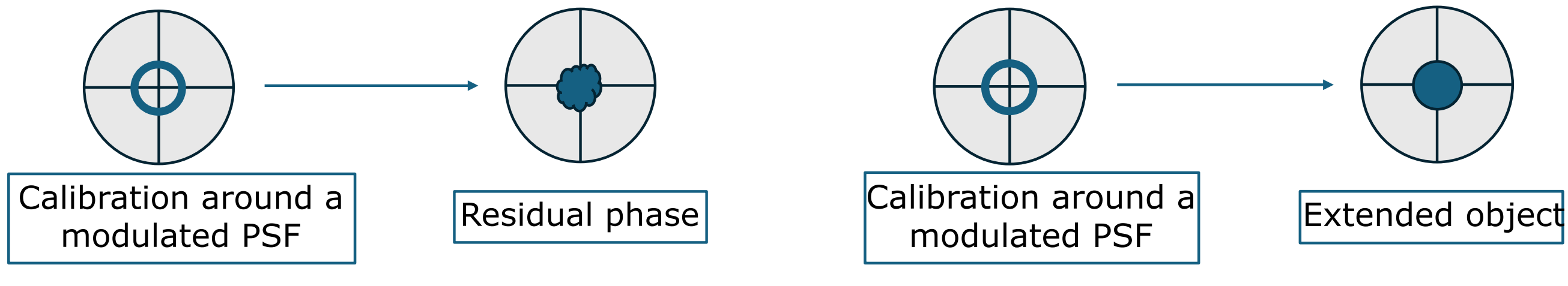

(a) Calibration for point sources. (b) Calibration for extended objects.

Figure 6: Calibration strategies for the non-modulated PWFS.

The core principle consists of recording the interaction matrix around a modulated PSF, while operating the closed-loop correction in a non-modulated regime. This provides the wavefront reconstructor with a structural prior, teaching the system to expect a spatially extended flux distribution rather than a perfect, diffraction-limited spot.

For point sources, on-sky closed-loop operations inevitably leave residual phase aberrations[23,27] that, combined with the internal vibrations of the bench, dynamically spread the PSF on the apex of the PWFS. Calibrating with a modulated PSF effectively anticipates this spread, significantly increasing the robustness of the non-modulated loop. Similarly, when tracking extended objects such as satellites, the spatial footprint of the target intrinsically spreads the light across the pyramid faces, acting as a natural optical modulation.[4,28] Therefore, the same modulated calibration strategy can be applied. The determination of the optimal modulation radius for calibration relative to the target's angular size is currently under investigation and will be detailed in an upcoming paper in preparation by Fétick et al.

In practice, this modulated calibration can be implemented either physically, classically by using a tip-tilt mirror to modulate the beam on the PWFS like in PAPYRUS, but in our case by driving the DM to apply a tip-tilt modulation during the interaction matrix recording, or numerically, by computing synthetic calibration matrices. The validity of this hybrid approach—modulated calibration coupled with non-modulated closed-loop operation—has been successfully demonstrated on the PAPYRUS bench. These results will be analyzed more precisely in an upcoming paper[29] but as a first analysis we can see that this strategy achieves Strehl ratio comparable to those of a classical, fully modulated PWFS on point sources (see Figure 7).

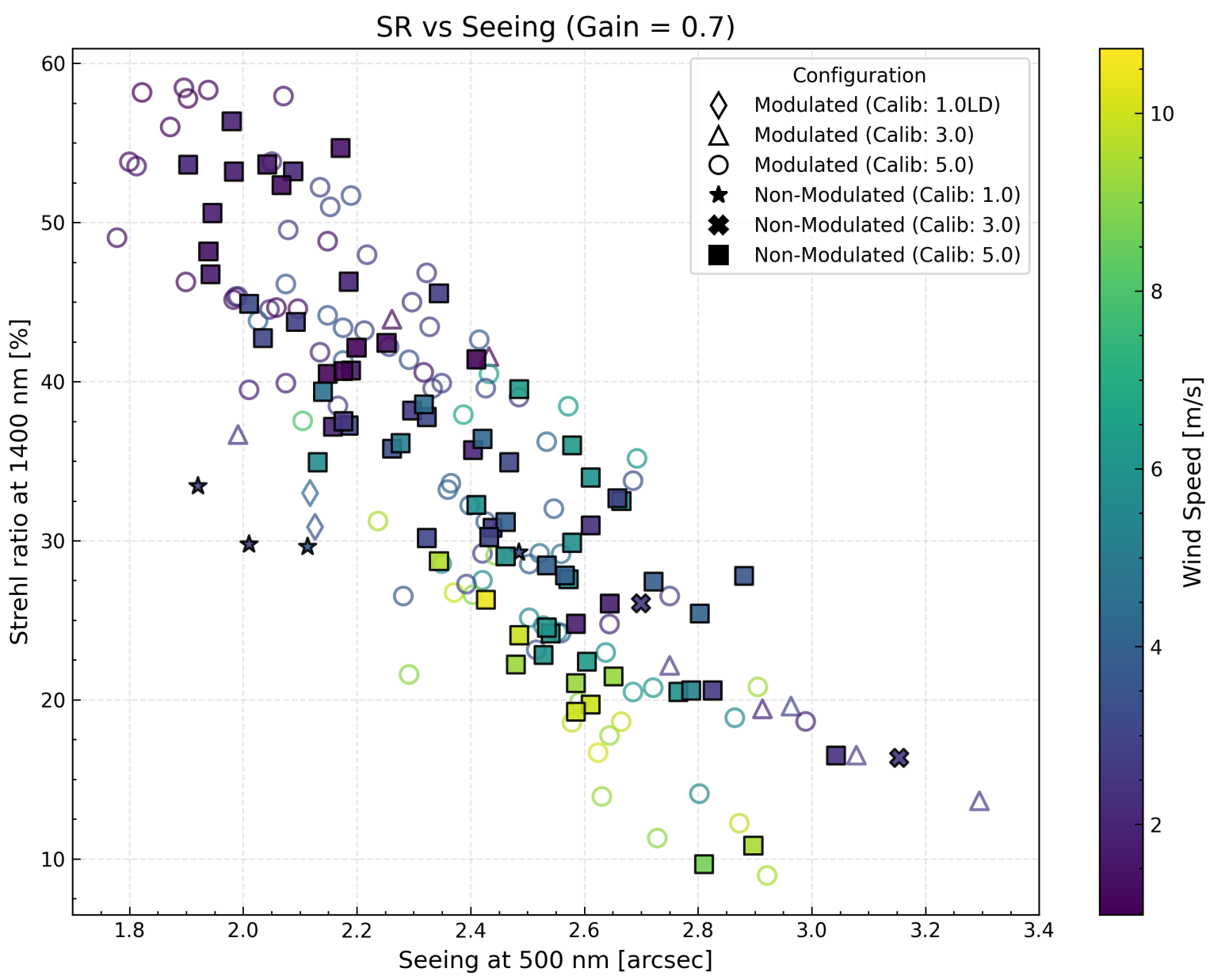

Figure 7: On-sky performance comparison from the PAPYRUS system. The non-modulated PWFS yields equivalent Strehl Ratios to the modulated configurations under various seeing.

# 3. PERFORMANCES SIMULATIONS

## 3.1 Simulation setup and atmospheric conditions

To predict the RAMA bench behavior and size the hardware parameters prior to on-sky operations, we explored the parameter space of the bench during the optical design part using AOPERA,[30,31] a Fourier bases AO simulator. The simulated atmospheric conditions cover the parameter space expected at the FEELINGS site. We simulate a 60 cm pupil with a Fried parameter ($r_0$) ranging from 4 cm to 10 cm defined at 500 nm, and equivalent wind speeds up to 10 m/s. The wavefront sensing and science evaluation are both computed at 635 nm. In parallel we also developed an End-To-End numerical twin using the OOPAO[32] simulation package.

Because satellite tracking involves low-elevation lines of sight,[1,2] we simulate zenith angles ranging from 0 to 70 degrees (elevations from 90°down to 20°). This extended optical path through the atmosphere significantly increases the scintillation, which is explicitly modeled in the simulations using the Angular Spectrum Method[33] to evaluate the PWFS robustness.

| **Parameter** | **Value** | **Unit** | **Remarks** |
|---|---|---|---|
| Telescope pupil diameter | 60 | cm | |
| Fried parameter ($r_0$) | 6 | cm (at 500 nm) | |
| Outer scale ($L_0$) | 25 | m | |
| Number of turbulent layers | 5 | – | |
| Layer altitudes | 0, 2, 5, 10, 15 | km | |
| Fractional $C_n^2$ weights | 30, 5, 5, 20, 40 | % | |
| Wind directions | 0, 72, 144, 216, 288 | ° | |
| Wavelength | 650 | nm | WFS and PSF |
| Synthetic modulation radius | 0 to 5 | $\lambda$/D | PWFS calibration |
| Zenith angles | 0–70 | ° | Elevations: 90°–20° |
| Loop frequencies | 500, 1000, 2000 | Hz | |
| Flux conditions | 0 – 9 | Magnitude (Visible) | |

Table 3: Simulation parameters for the parametric study.

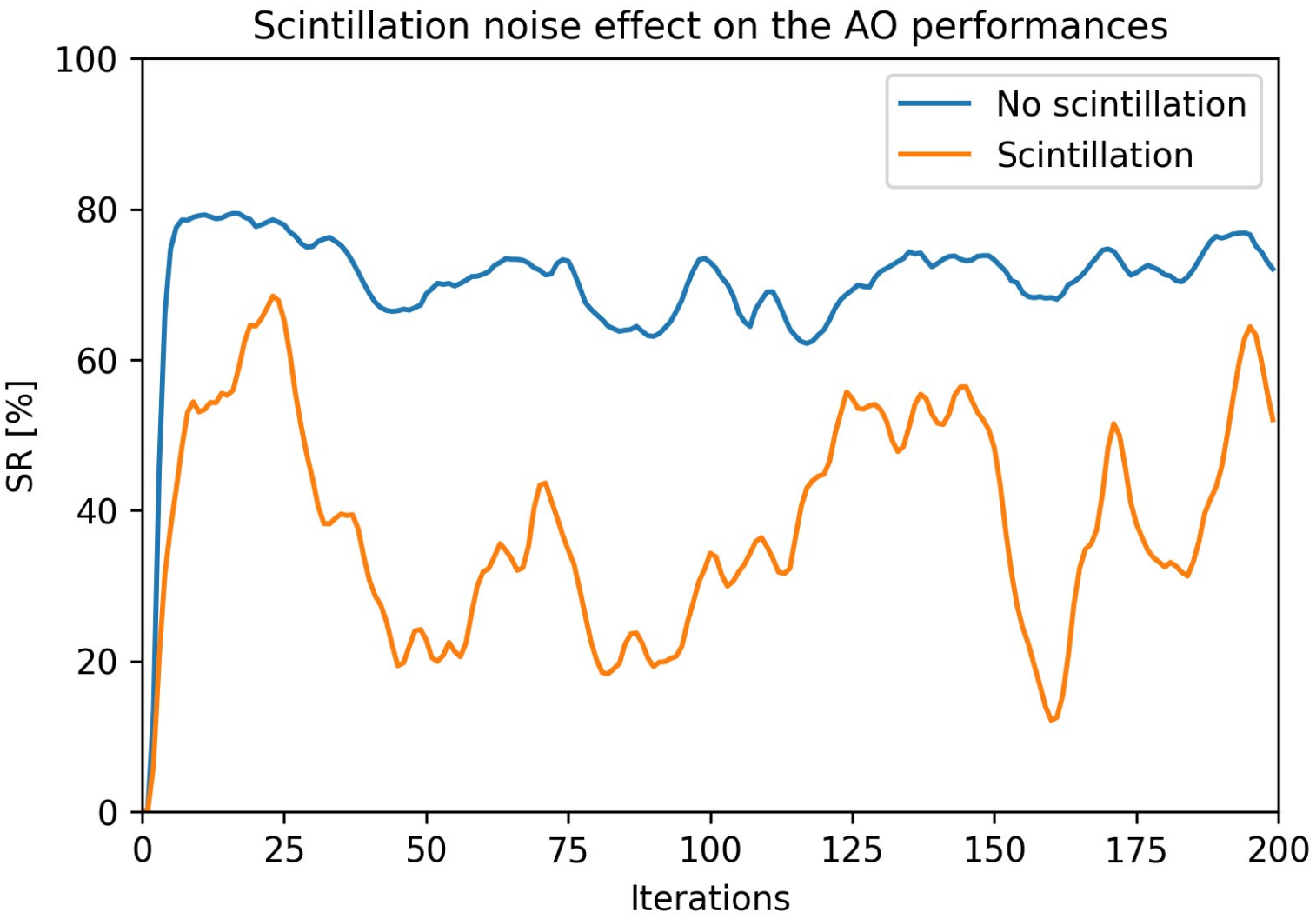

Figure 8: Impact of scintillation on the Strehl Ratio (OOPAO simulations with/without scintillation, 200 frames, Rytov $\approx 0.8$).

The precise behavior of the PWFS in a scintillation regime and what can be gained from the scintillation information[34,35]will be more thoroughly characterized in a following paper,[29] but first simulations are showing us as expected that the scintillation noise is degrading the close loop performances using a PWFS in a strong scintillation regime with a rytov of around 0.8 (see Figure 8).

### 3.2 Error budget analysis

This error budget analysis, conducted using AOPERA across varying elevations, guide star magnitudes, and loop frequencies, served a dual purpose. Initially, we used it to constrain the adaptive optics design of the RAMA testbench by highlighting its operational limits and identifying critical areas requiring specific attention. Now that the hardware baseline is fixed, this study allows us to estimate the expected performances and system boundaries.

Specifically, with two distinct DMs integrated into the bench, a key objective is to determine the optimal observational conditions for each. For this initial parametric study, the simulated system operates in a standard Single Conjugate Adaptive Optics (SCAO) configuration using only one DM at a time. Complex control architectures, such as a woofer-tweeter configuration commanding both the DM97 and DM192 simultaneously, will be addressed in future studies.

Figure 9 and Figure 10 illustrate the variance breakdown as a function of the elevation for the DM97 and DM192 configurations, respectively. To frame the operational limits of the RAMA testbench, two distinct tracking regimes were evaluated: a high-flux scenario (Mag 0.0 operating at 2000 Hz) and a fainter target scenario (Mag 4.0 operating at 1000 Hz).

The variance decomposition reveals two distinct behaviors depending on the flux:

- **In the high-flux regime (Mag 0.0):** The system is limited by the temporal and fitting errors. As the elevation drops toward 20°, the fast apparent wind speeds cause the temporal variance to drastically increase, accounting for almost half of the total error budget. This is accompanied by a noticeable rise in scintillation, while aliasing and chromatism remain marginal. In this regime, the DM192 naturally has a lower fitting error compared to the DM97.
- **In the low-flux regime (Mag 4.0):** The wavefront sensing noise—specifically the Read-Out Noise (RON) from the IMX426 camera and photon noise—becomes the dominant limiting factor, except at low elevation where the temporal error starts being impactful as well.

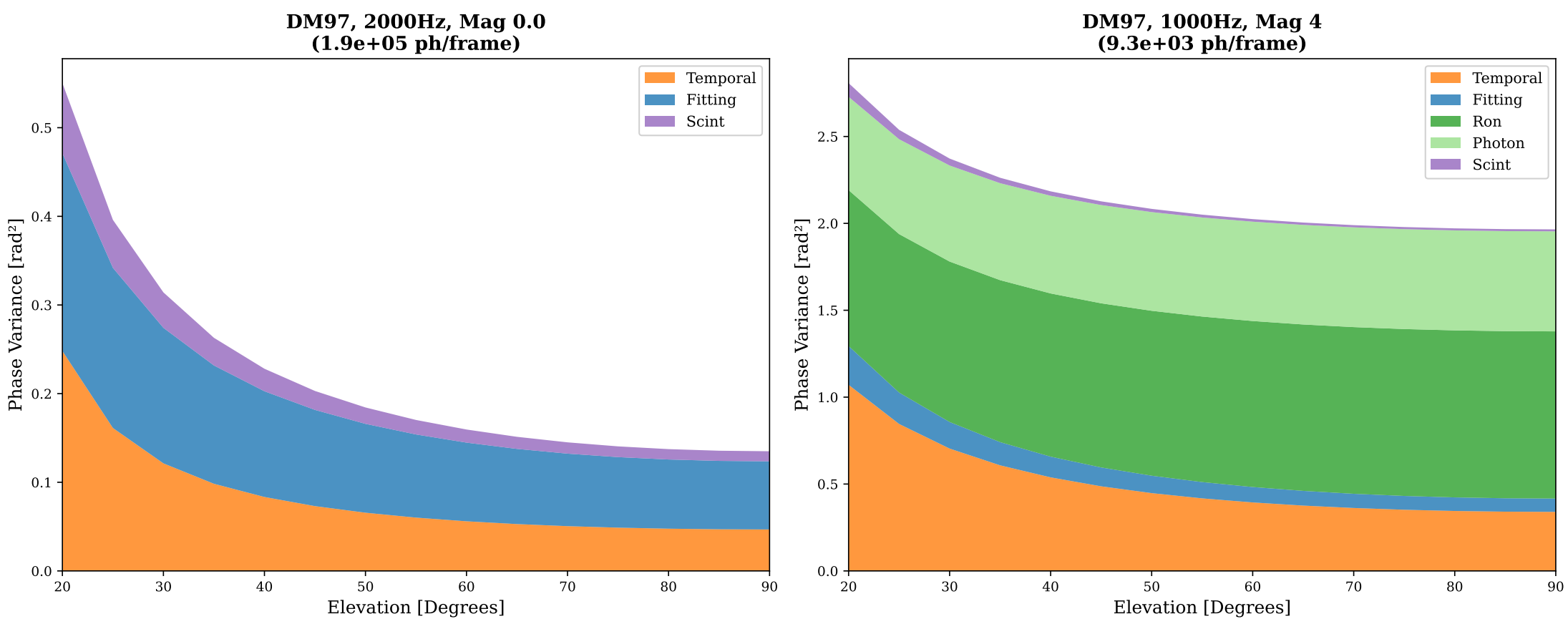

Figure 9: Simulated error budget breakdown as a function of elevation for the DM97 configuration. Left: High-flux regime (Mag 0.0 at 2000 Hz) dominated by temporal and fitting errors. Right: Low-flux regime (Mag 4.0 at 1000 Hz) dominated by Read-Out Noise (RON) and photon noise.

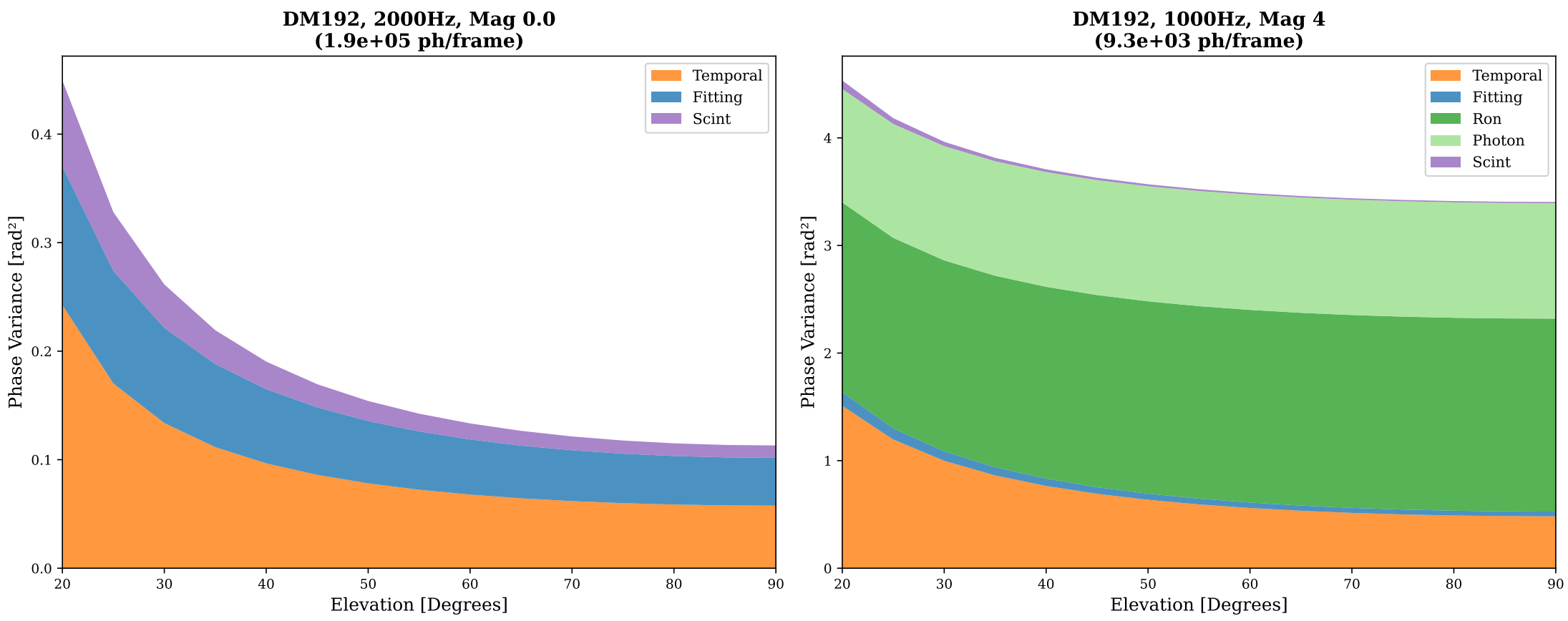

Figure 10: Simulated error budget breakdown as a function of elevation for the DM192 configuration under the same conditions. The higher actuator density amplifies the noise propagation in the low-flux regime (Right) compared to the DM97.

### 3.3 Trade-off study: actuator density vs. loop frequency

To further evaluate the expected performances of the bench, we evaluated both DM geometries across different loop frequencies (500 Hz, 1000 Hz, and 2000 Hz) to map the optimal operational regimes.

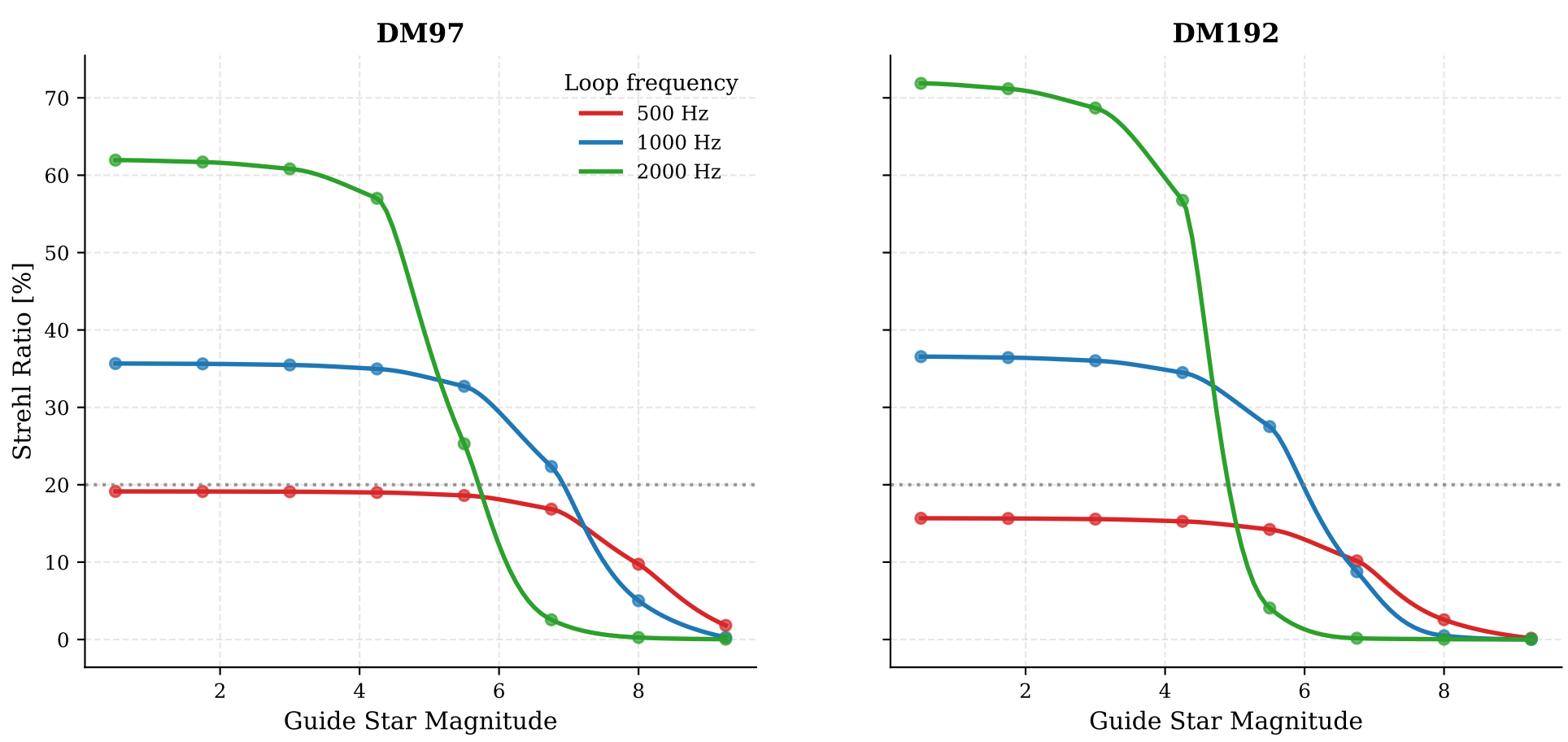

Figure 11: Strehl Ratio versus guide star magnitude for the DM97 and DM192 configurations evaluated at a fixed elevation of 45°.

As highlighted in the previous error budget analysis, increasing the number of actuators does not systematically give better performance. Controlling 192 actuators increases the degrees of freedom in the reconstructor. For a given total flux, this divides the incoming light into smaller sub-apertures, making the inverse problem less well-posed. This amplifies the propagation of the Read-Out Noise (RON) into the phase variance (as seen when comparing the right panels of Figures 9 and 10).

At low loop frequencies or in low flux regimes, the DM97 configuration therefore significantly outperforms the DM192 configuration thanks to its better Signal-to-Noise Ratio (SNR) per sub-aperture. The DM192 configuration only becomes advantageous at high flux levels and high loop frequencies ($\geq$ 1000 Hz), where the noise is negligible and the temporal error is sufficiently contained to benefit from the reduced fitting error.To precisely quantify this transition, we simulated the Strehl Ratio versus guide star magnitude (see Figure 11). The results show that tracking faint targets requires operating the DM97 to preserve a stable loop, while the DM192 performs better under high flux conditions.

# 4. CURRENT STATUS AND FIRST RESULTS

## 4.1 Prototype bench

Prior to the on-sky deployment, we assembled and aligned a complete prototype of the RAMA testbench in the laboratory using the final hardware (RTC, cameras, DM, Pyramid) (see Figure 12). This setup includes a dedicated telescope simulator, the common optical path, and the non-modulated PWFS branch. Interfacing this hardware with the DAO RTC provided a controlled environment to validate the physical optical design against numerical simulations, and to debug the overall software architecture (telemetry acquisition, phase reconstruction, and real-time control).

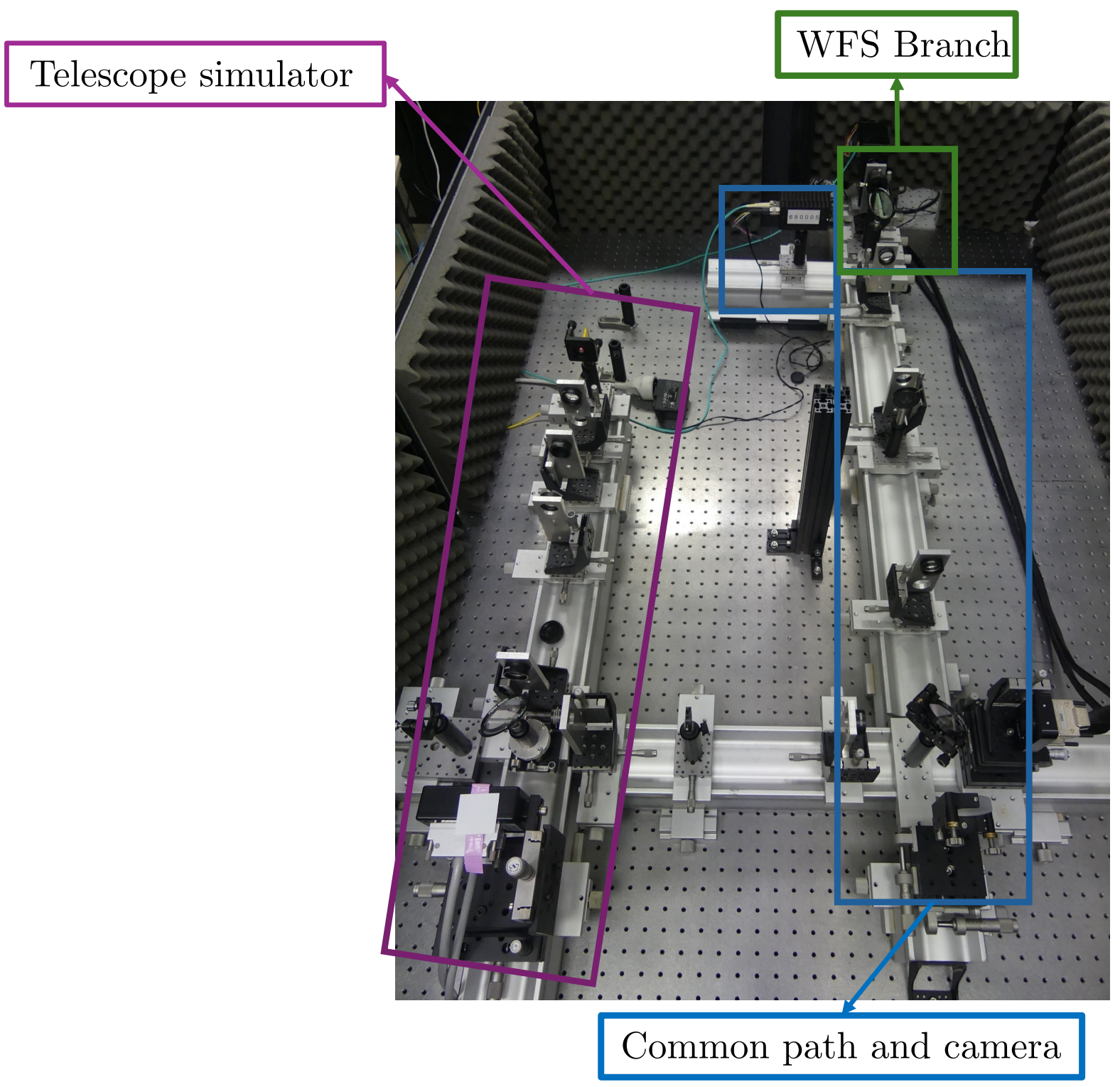

Figure 12: The RAMA prototype bench in the laboratory.

## 4.2 First closed-loop results on the prototype bench

Using the internal calibration source, we successfully closed the AO loop on this prototype. The frequency was temporarily limited to 100 Hz by the current DM electronic, which will be upgraded for the final instrument.

The modularity of the DAO RTC facilitated the rapid implementation and swapping of different control algorithms. As shown in Figures 13 and 14, we conducted initial tests to benchmark a standard linear reconstructor coupled with a classical integrator against advanced data-driven methods. These tests confirm that the non-modulated PWFS can effectively drive the loop using both classical and neural network-based approaches.

Specifically, we demonstrated that Convolutional Neural Networks (CNNs) are now fully functional within our real-time architecture to handle the PWFS non-linearities, while Reinforcement Learning (RL) agents successfully manage the loop control. Validating these algorithms on the hardware will allow us to simultaneously test them on-sky to mitigate two major error sources expected: wavefront sensing non-linearities and temporal lag.

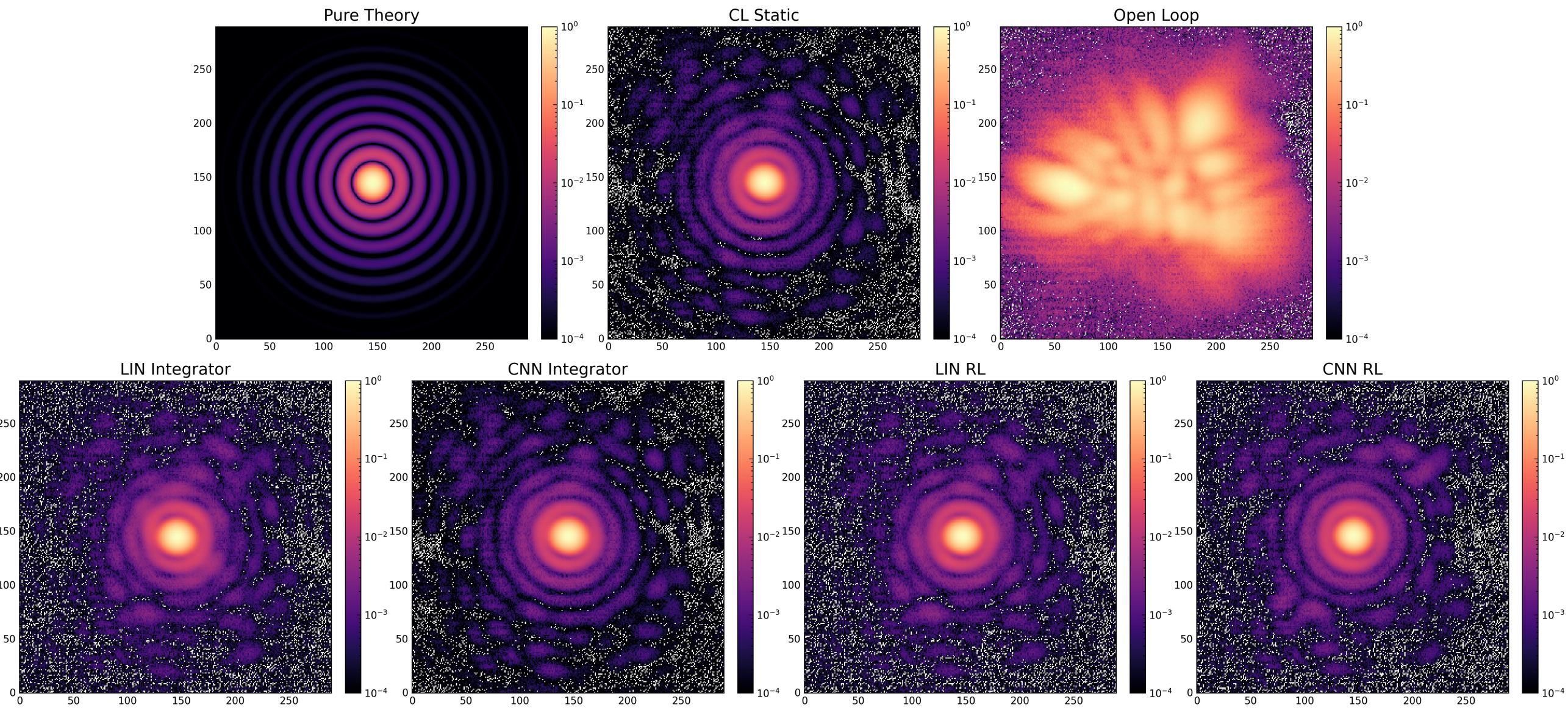

Figure 13: Experimental PSF comparison obtained on the RAMA prototype bench (colorbar in log scale).

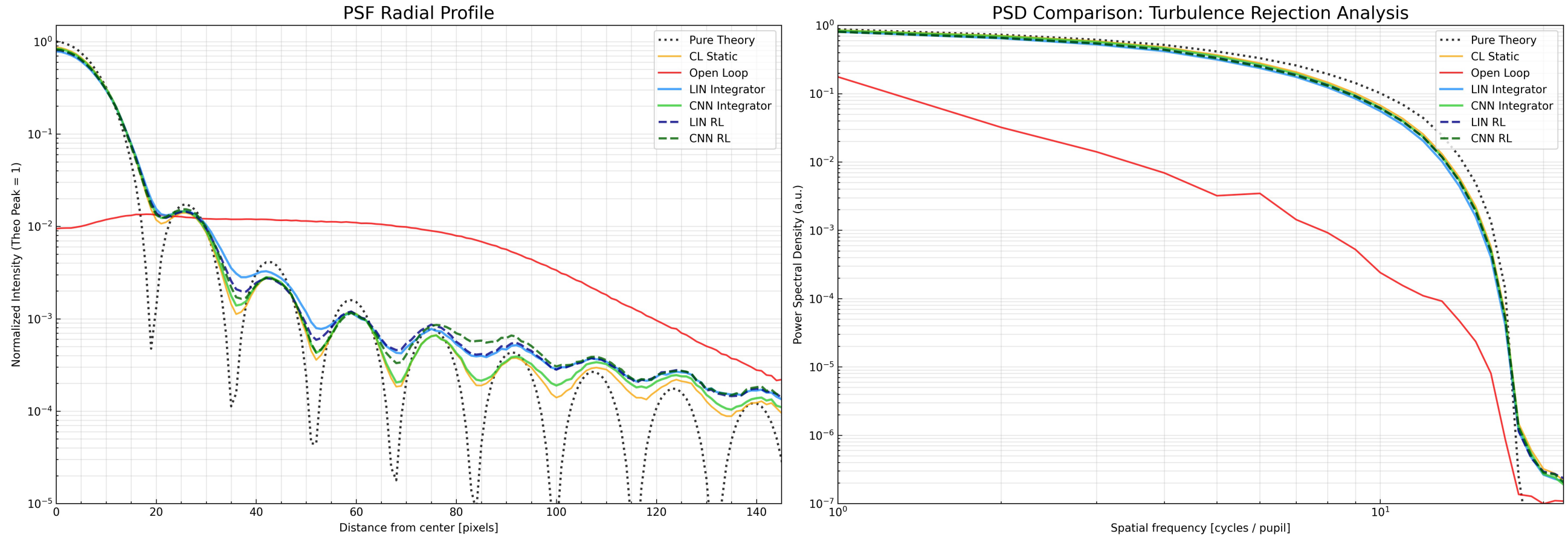

Figure 14: Power Spectral Density (PSD) and PSF radial profile comparison.

### 4.3 Telescope integration and first on-sky commissioning

Following the successful validation of the hardware and software interfaces in the laboratory, the final opto-mechanical structure was assembled. The mechanical interfaces designed to mount the bench at the Nasmyth port of the FEELINGS telescope were manufactured, and the optical breadboard has been successfully installed in the telescope dome.

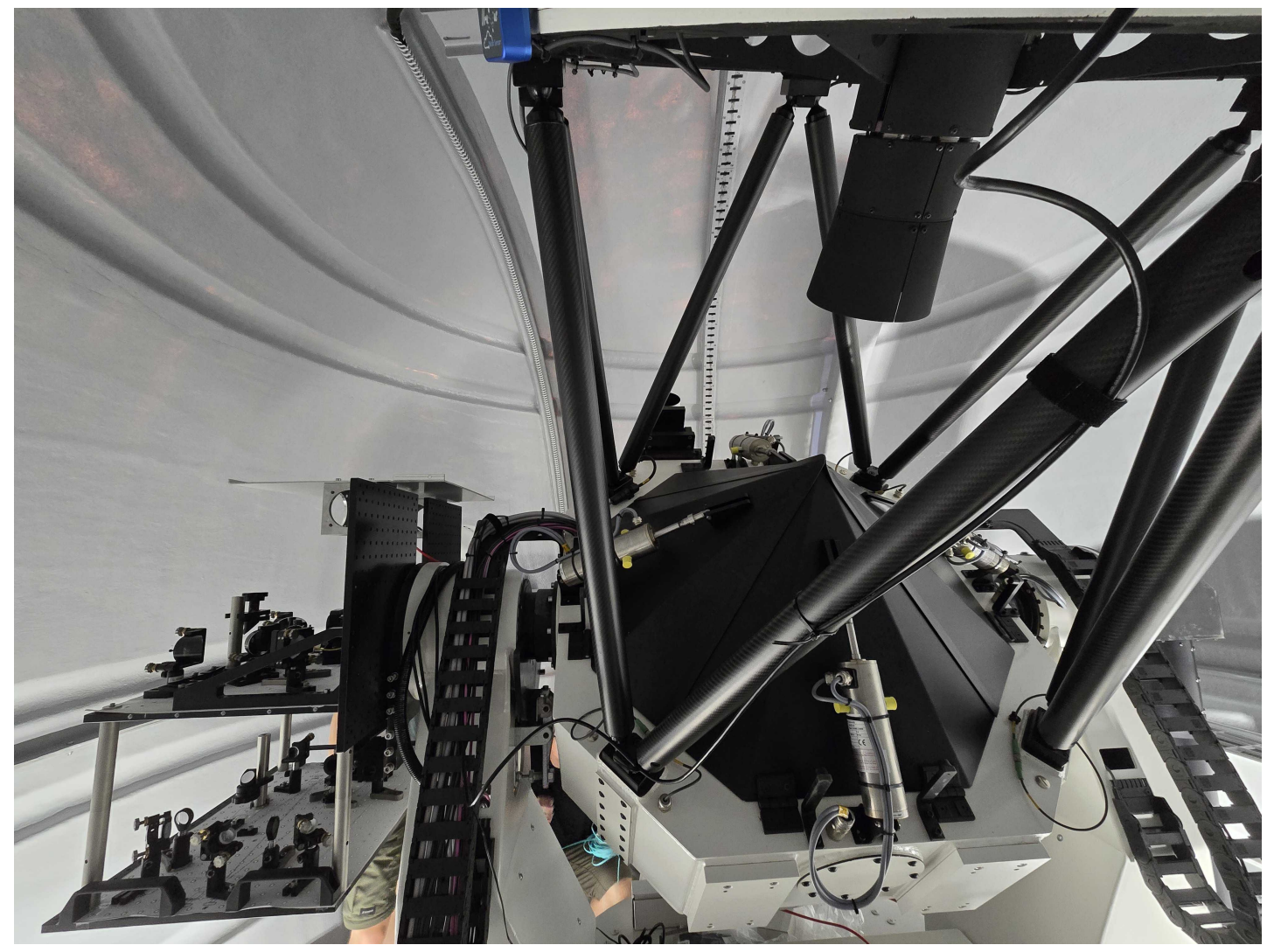

(a) RAMA optical bench (left of the picture) integrated at the telescope Nasmyth port.

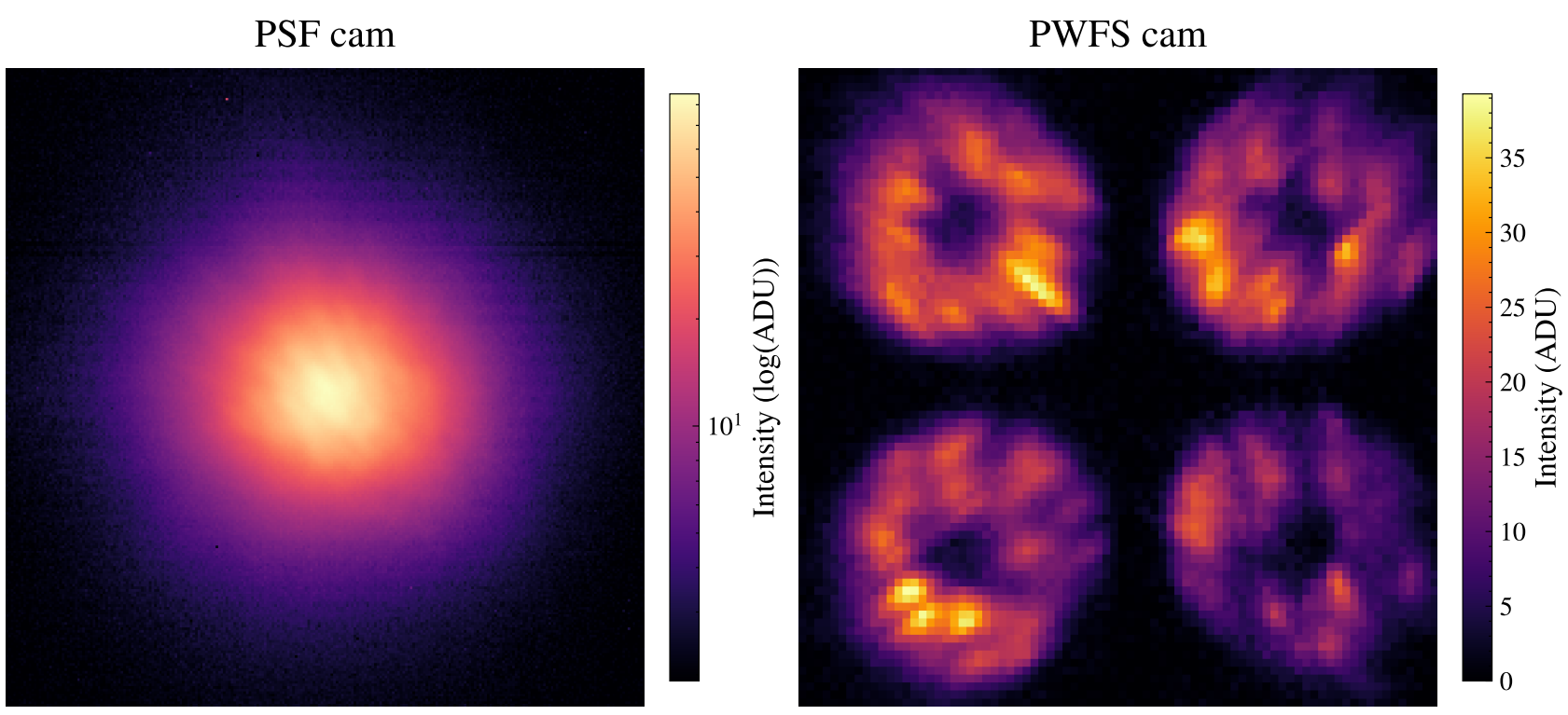

(b) First light data of the PSF camera (left) and PWFS camera (right).

Figure 15: Successful on-sky integration and first open-loop validation of the RAMA testbench.

The system recently achieved its first on-sky technical light during two initial commissioning days and nights (15-16th of July 2026). During those two days, we managed to take the bench from the laboratory in Marseille, mount it at the telescope (see Figure 15a) near Toulouse, check the internal alignment on a calibration source, and align the bench with the telescope to get the first light on the two cameras (see Figure 15b). These preliminary observations were strictly dedicated to validating the opto-mechanical coupling between the telescope and the bench, verifying the internal alignment on actual astronomical sources, and confirming the stability of the closed-loop control system under real atmospheric conditions.

Although the primary goal of these initial sessions was not to conduct comprehensive performance evaluations, the observed nominal system behavior provides a very encouraging starting point. The bench is now permanently on-sky and almost fully controllable remotely, greatly facilitating future operations. Our immediate next steps will involve performance characterization nights to evaluate the system's absolute correction capacities and optimize the Strehl ratio. In parallel, planned hardware upgrades will be integrated, including the addition of a tracking camera, a light curve camera, and a second DM. Once this baseline AO performance is validated and the final hardware is in place, future observation campaigns will focus on our main scientific objectives: demonstrating non-modulated calibration strategies on real extended objects and tracking low-elevation satellites under strong scintillation conditions.

## 5. CONCLUSION

The RAMA testbench provides an experimental platform to address the specific challenges of extended-object observations, specifically satellite tracking. Placed at the Nasmyth focus of the 60 cm FEELINGS telescope, the system is designed to test and validate wavefront control strategies under strong scintillation and on extended targets in preparation for the upcoming PROVIDENCE ground station.

The system baseline relies on a visible, non-modulated Pyramid Wavefront Sensor. The optical design was optimized to match the specific geometry of the pyramid's apex, operating at a high focal ratio to restrict the rooftop footprint to 25% of the PSF size. Numerical simulations mapped the error budget, confirming that the temporal error acts as the dominant limitation for low-elevation satellite tracking. This parametric study also established the operational limits of the hardware, demonstrating that operating the DM97 is preferable at low fluxes or low loop frequencies, while the DM192 becomes advantageous at high loop frequencies and high fluxes.

RAMA supports the integration of advanced control laws. The laboratory prototype has been successfully assembled and interfaced with the DAO RTC. We have demonstrated the ability to close the AO loop on the bench at 100 Hz (prototype) with improvements expected on the final bench (500-2000 Hz), validating both classical control pipelines (Linear reconstructor, Integrator) and data-driven methods (CNN, Reinforcement Learning).

Beyond the technical validation of the components, a major outcome of this project is demonstrating the ability to rapidly transition from concept and simulation to an on-sky operational testbench for small AO systems like ours. The entire process—encompassing optical and mechanical design, laboratory integration, telescope mounting, and first technical light—was successfully achieved in just 4 to 5 months. With the testbench now on-sky and largely remotely operable, upcoming steps will focus on validating AO performances and integrating the final hardware components. These milestones will ultimately pave the way for our primary scientific campaigns: evaluating non-modulated PWFS performance during real satellite observations and validating the robustness of data-driven algorithms under actual atmospheric conditions.

## ACKNOWLEDGMENTS

This work benefited from the support the French National Research Agency (ANR) with the Programme Investissement Avenir F-CELT (ANR-21-ESRE-0008), PEPR ORIGINS (XAO-WFS ANR-22-EXOR-00062 and COMPACT SPECTROGRAPH ANR-22-EXOR-0006), the ANR-DGA-AID ASTRID program (ANR-25-ASTR-0015), the CNRS-IRP GALILEAO program, the Action Spécifique Haute Résolution Angulaire (ASHRA) of CNRS/INSU co-funded by CNES, the french government under the France 2030 investment plan (cassiopée project) and the Initiative d'Excellence d'Aix-Marseille Université A*MIDEX, program number AMX-22-RE-AB-151. PROVIDENCE is co-funded by the Région Sud Provence-Alpes-Côte d'Azur and the European Union, and is developed by ONERA. It is installed at Observatoire de Haute-Provence. PROVIDENCE is supported by the European Regional Development Fund (ERDF), under the Provence-Alpes-Côte d'Azur Region and Alps Massif / ERDF-ESF+JTF 2021-2027 program, to support projects contributing to the economic, scientific, and technological development of the territory.